\documentclass[aps,prb,amsmath,amssymb,twocolumn,floatfix]{revtex4-1}
\usepackage{graphicx}
\usepackage{dcolumn}
\usepackage{bm}

\begin{document}

\title{Freezing out of a low-energy bulk spin exciton in SmB$_6$}

\author{K. Akintola,$^1$ A. Pal,$^1$ S. R. Dunsiger,$^1$ A. C. Y. Fang,$^1$ M. Potma,$^{1,2}$ S. R. Saha,$^3$ X. F. Wang,$^3$ J. Paglione,$^{3,4}$ and J. E. Sonier,$^{1,4}$}

\affiliation{$^1$Department of Physics, Simon Fraser University, Burnaby, British Columbia V5A 1S6, Canada \\
$^2$Kwantlen Polytechnic University, Richmond, British Columbia V6X 3X7, Canada \\
$^3$Center for Nanophysics and Advanced Materials, Department of Physics, University of Maryland, College Park, Maryland 20742, USA \\
$^4$Canadian Institute for Advanced Research, Toronto, Ontario M5G 1Z8, Canada}

\date{\today}
\begin{abstract}
The Kondo insulator SmB$_6$ is purported to develop into a robust topological insulator at low temperature. 
Yet there are several puzzling and unexplained physical properties of the insulating bulk. 
It has been proposed that bulk spin excitons may be the source of these anomalies and may also adversely 
affect the topologically-protected metallic surface states. Here, we report muon spin rotation ($\mu$SR)
measurements of SmB$_6$ that show thermally-activated behavior for the temperature dependences of the
transverse-field (TF) relaxation rate below 20~K and muon ($\mu^+$) Knight shift below 5-6~K. Our data are 
consistent with the freezing out of a bulk low-energy ($\sim \! 1$~meV) spin exciton
concurrent with the appearance of metallic surface conductivity. Furthermore, our results support the
idea that spin excitons play some role in the anomalous low-temperature bulk properties of SmB$_6$.
\end{abstract}

\maketitle
\noindent {\bf INTRODUCTION} \\
Due to a combination of spin-orbit coupling and time reversal symmetry, a topological insulator (TI) supports protected metallic edge and surface states 
in two-dimensional (2-D) and three-dimensional (3-D) systems, respectively.\cite{Hasan:10} The ideal 3-D TI has a truly insulating bulk gap,
as this restricts applications of the transport properties to the topologically-protected surface, where the electron spin is uniquely locked 
to the charge momentum. Yet true bulk insulating behavior is not realized in established TIs due to bulk impurity conduction.\cite{Ando:13}
The homogeneous intermediate-valence compound SmB$_6$ is a strong candidate for a 3-D TI with a robust bulk insulating gap.\cite{Dzero:16}
In contrast to a conventional band insulator, the insulating gap in SmB$_6$ is created
via Kondo hybridization of localized Sm-$4f$ and itinerant Sm-$5d$ electrons, with the Fermi level residing in the hybridization gap. 
  
Experimental evidence for SmB$_6$ being a TI is provided by transport measurements that have demonstrated 
predominant surface electrical conduction below 5-7~K,\cite{Kim:13,Wolgast:13,Kim:14} and the detection of in-gap surface states by 
angle-resolved photoemission spectroscopy (ARPES).\cite{Jiang:13,Xu:13,Neupane:13,Frantzeskakis:13,Xu:14a}
However, recent high-resolution ARPES results suggest that the surface conductivity is not associated with toplogical surface states.\cite{Hlawenka:18}
Moreover, at low $T$ there is a sizable metallic-like linear-$T$ specific heat of bulk origin,\cite{Wakeham:16} and significant 
bulk ac-conduction.\cite{Laurita:16} Quantum oscillations are observed in the 
magnetization of SmB$_6$ as expected for 2-D metallic surface states,\cite{Li:14} but subsequent measurements suggest the origin is
a bulk 3-D Fermi surface.\cite{Tan:15} These findings have raised the possibility of charge-neutral 
fermions in the insulating bulk.\cite{Baskaran:15,Erten:17,Chowdhury:17}

The Sm ions in SmB$_6$ rapidly fluctuate between non-magnetic Sm$^{2+}$ ($4f^6$) and magnetic Sm$^{3+}$ (4f$^55d^1$) 
electronic configurations, resulting in an average intermediate valence that varies with temperature.\cite{Tarascon:80,Mizumaki:09}
Interestingly, SmB$_6$ exhibits magnetic fluctuations below 20-25~K where the Kondo gap is fully formed, as observed by muon spin relaxation/rotation
($\mu$SR).\cite{Biswas:14,Akintola:17} This was first presumed\cite{Biswas:14} to be due to the bulk magnetic in-gap states detected by
nuclear magnetic resonance (NMR) below 20~K,\cite{Caldwell:07} and later specifically speculated to be due to
bulk spin excitonic excitations.\cite{Biswas:17} Spin excitons in SmB$_6$ are induced by residual dynamic AFM exchange 
interactions between the hybridized quasiparticles and are a precursor to an AFM instability.\cite{Riseborough:00,Riseborough:03}
A 14~meV bulk collective mode observed within the hybridization gap by inelastic neutron scattering (INS) has been 
interpreted as a spin exciton.\cite{Alekseev:95,Fuhrman:15} Bulk spin excitons are expected to adversely affect
the protected topological order by causing spin-flip scattering of the surface states.\cite{Kapilevich:15}
There is some indirect evidence for this from recent angle-integrated photoemission\cite{Arab:16} and 
planar tunneling\cite{Park:16} spectroscopy studies.
 
Since an implanted $\mu^+$ does not create a spin-exciton excitation, $\mu$SR is only sensitive to thermally-activated spin excitons.
Consequently, the 14 meV bulk spin exciton observed by INS should not be detectable by $\mu$SR below 20-25~K. 
An additional lower energy ($\lesssim \! 1$~meV) spin-exciton branch has recently been predicted and
suggested to contribute to bulk quantum oscillations and cause the anomalous upturn in the specific heat at low $T$.\cite{Knolle:17}
On the other hand, the low-$T$ specific heat is enhanced by Gd impurities\cite{Fuhrman:17} and drastically reduced in isotopically enriched 
SmB$_6$,\cite{Orendac:17} suggesting magnetic impurities play some role but spin excitons do not. Nevertheless,
decoupling of the surface states from a $\sim \! 4$~meV bulk spin exciton has been argued to explain the rapid increase in the 
surface conductance below 5-6~K and subsequent saturation below 4~K.\cite{Park:16} While the reduction in energy from 14~meV to 4~meV 
is assumed to be due to a diminished Kondo temperature at the surface,\cite{Kapilevich:15} a distinct low-energy bulk 
spin exciton is another possibility.
      
Here we report high TF-$\mu$SR measurements of the $\mu^+$-Knight shift in Al-flux grown SmB$_6$ 
single crystals that provide evidence for a bulk spin exciton of energy much lower than 14~meV.  
Our data suggest that the occurrence of a resistivity plateau below $T \! \sim \! 4$~K is associated with the
freezing out of an $\sim \! 1$~meV spin exciton, such that metallic surface states emerge when spin exciton 
scattering becomes negligible. \\ 

\begin{figure}
\centering
\includegraphics[width=8.0cm]{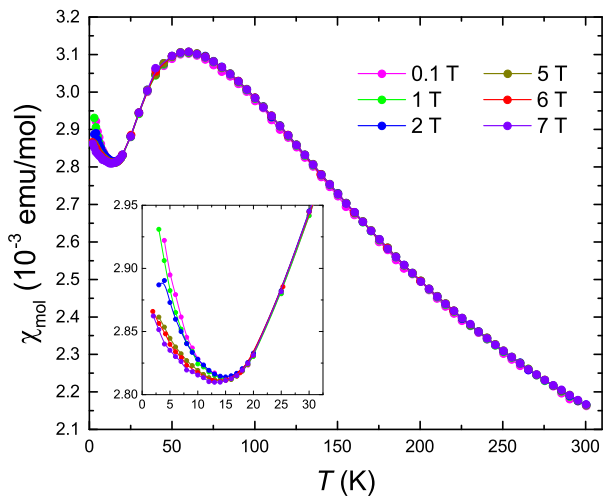}
\caption{Temperature dependence of the bulk magnetic susceptibility for fields applied parallel to the $c$-axis.
The inset is a blow up of the low-$T$ data.}
\label{fig1}
\end{figure}

\noindent {\bf RESULTS} \\
Figure~\ref{fig1} shows the temperature dependence of the bulk magnetic susceptibility $\chi_{\rm mol}$ at different magnetic fields {\bf H} applied
parallel to the $c$-axis. At high temperature $\chi_{\rm mol}$ exhibits Curie-Weiss behavior indicative of paramagnetic Sm ions.   
Opening of the hybridization gap below 110~K gives rise to the broad maximum, followed by a field-dependent upturn below 
$T \! \sim \! 17$~K that masks the expected low-$T$ van-Vleck saturation. The upturn has previously been 
attributed to impurities.\cite{Roman:96,Gabani:02} 

In the absence of spontaneous magnetic order, the applied field polarizes the conduction electrons and induces 
spin polarization of the localized Sm-$4f$ magnetic moments. The local field {\bf B}$_\mu$ sensed by the $\mu^+$ is the vector sum of the
dipolar field ${\bf B}_{\rm dip}$ from the Sm-$4f$ magnetic moments and a contact hyperfine field ${\bf B}_{\rm c}$. 
At high $T$ where SmB$_6$ behaves as a poor metal, the muon's positive charge is screened by a cloud of conduction electrons. The screening electron cloud
acquires a finite spin density due to the Ruderman-Kittel-Kasuya-Yosida (RKKY) interaction with the spin-polarized Sm-$4f$ moments,
and by direct contact generates a hyperfine field at the $\mu^+$ site. This is expected to vanish with the development
of a bulk insulating gap at lower $T$. In an insulating state ${\bf B}_{\rm c}$ may instead originate from direct overlap of 
the $\mu^+$ with the wavefunction of localized magnetic electrons, or from bonding of the $\mu^+$ to an ion that is covalently bonded 
to a local atomic magnetic moment.\cite{Schenck:85}

\begin{figure}
\centering
\includegraphics[width=9.0cm]{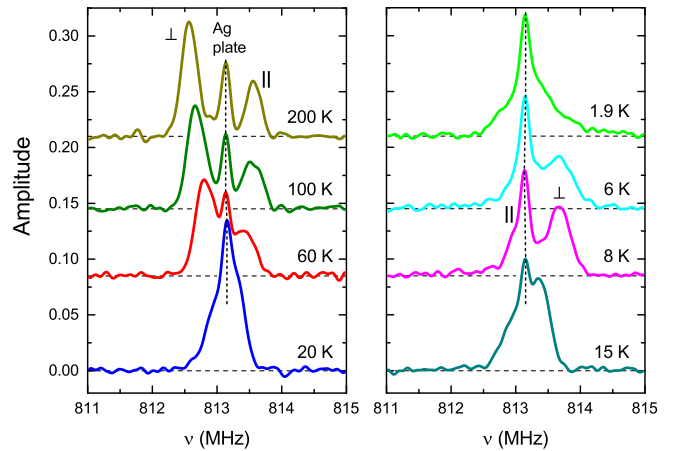}
\caption{Fourier transforms of representative TF-$\mu$SR spectra at $H \! = \! 6$~T applied parallel to the $c$-axis. 
The frequency $\nu$ is equivalent to $(\gamma_\mu/2 \pi)B_\mu$.}
\label{fig2}
\end{figure}

Figure~\ref{fig2} shows Fourier transforms of TF-$\mu$SR time spectra recorded on SmB$_6$ at $H \! = \! 6$~T. Due to
the apodization necessary to remove ringing artifacts caused by the short muon time window ($\sim \! 10$~$\mu$s) 
and noise caused by fewer counts at later times (due to the short muon lifetime), the Fourier transforms 
are a broadened visual approximation of the internal magnetic field distribution. Consequently, analysis of the 
TF-$\mu$SR signals were done in the time domain. At $T \! =  \! 200$~K there are three well separated peaks 
in the Fourier transform. The central peak arises from muons stopping in the 
Ag backing plate. The left and right peaks have an amplitude ratio of 2:1, and are consistent with 
the $\mu^+$ stopping at the midpoint of the horizontal or vertical edges of the cubic Sm-ion sublattice.
This is in agreement with the identified $\mu^+$ site in CeB$_6$.\cite{Schenck:02} Moreover, we have verified
the $\mu^+$ site assignment by TF-$\mu$SR measurements with {\bf H} applied $45^{\circ}$ with respect to the $c$-axis (see Fig.~S3).   

\begin{figure}
\centering
\includegraphics[width=9.0cm]{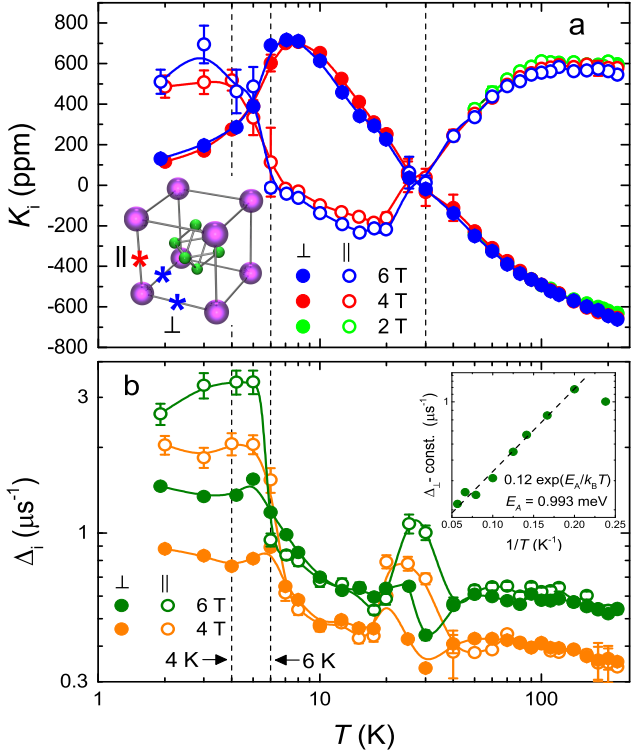}
\caption{Temperature dependence of {\bf a} the $\mu^+$-Knight shifts, and {\bf b} the TF-$\mu$SR relaxation rates at magnetic fields of 
4~T and 6~T applied parallel to the $c$-axis. Data at $H \! = \! 2$~T is also shown in {\bf a} for $T \! \geq \! 50$~K, below which the 
different components of the TF-$\mu$SR signal are not clearly resolved. The error bars represent the uncertainties in the parameters
$\nu_i$ and $\Delta_i$ from the fits in the time domain.    
The inset in {\bf a} shows the two magnetically-inequivalent muon sites on the vertical ($\parallel$) and 
horizontal edges ($\perp$) of the Sm-ion cubic sublattice. The inset in {\bf b} shows a semi-log plot of
$\Delta_{\perp}$ (minus a constant) versus $1/T$, for $4.2 \! \leq \! T \! \leq \! 17.6$~K. The straight dashed line is a fit
of the data for $T \! \geq \! 5$~K to a thermally-activated function.}
\label{fig3}
\end{figure}

\begin{figure}
\centering
\includegraphics[width=9.0cm]{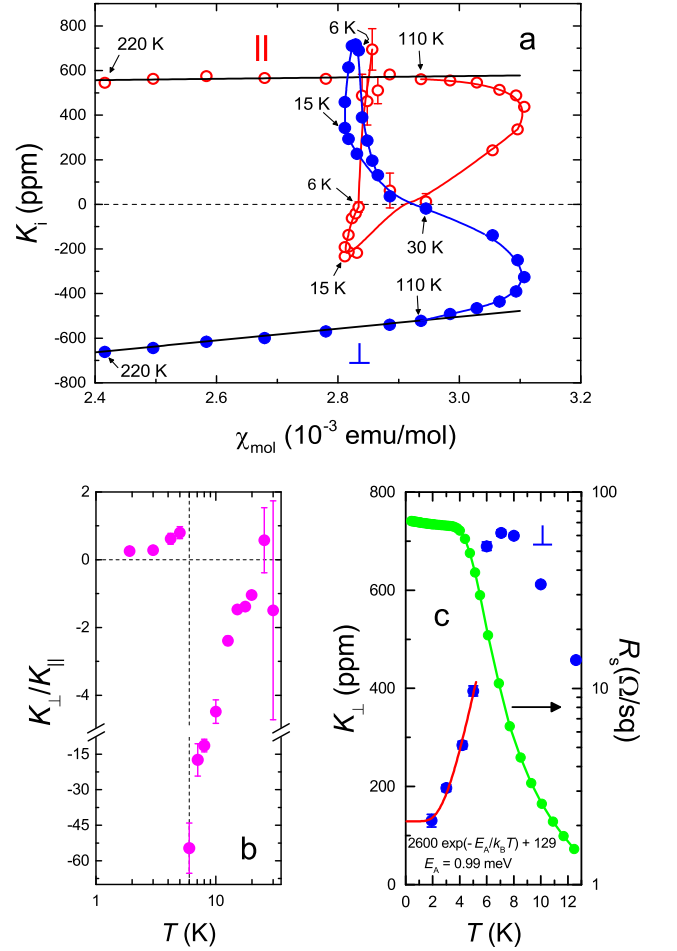}
\caption{{\bf a} Knight shift at the two magnetically-inequivalent $\mu^+$ sites versus $\chi_{\rm mol}$ at $H \! = \! 6$~T. 
Temperature is an implicit parameter. The straight black lines are fits to equations (\ref{eqn:Kperp}) and (\ref{eqn:Kpara}).
Temperature dependence of {\bf b} the ratio $K_{\perp}/K_{\parallel}$ below 30~K, and  
{\bf c} $K_{\perp}$ and the electrical sheet resistance below 13~K. The red curve is a fit of the $K_{\perp}$ data
for $T \! \leq \! 5$~K to a thermally-activated Arrhenius equation, assuming an activation energy $E_{\rm A} \! = \! 0.99$~meV.}
\label{fig4}
\end{figure}

In a field applied parallel to the $c$-axis, the $\mu^+$ site is magnetically inequivalent on
the horizontal and vertical edges of the cubic Sm sublattice. The dipole field generated by polarization of the Sm-$4f$ moments along the 
{\bf c} direction is equivalent and antiparallel to {\bf H} at the ($\perp$) sites ($\frac{1}{2}$, 0, 0) and (0, $\frac{1}{2}$, 0), and
different in magnitude and parallel to {\bf H} at the ($\parallel$) site (0, 0, $\frac{1}{2}$) (see Fig.~\ref{fig3}a inset). Consequently,
the TF-$\mu$SR time spectra were fit to an asymmetry function with a two-component sample contribution (see Fig.~S4)
\begin{eqnarray}
A(t) & = & A_{\rm s} \left[\frac{2}{3}e^{-\Delta_\perp^2 t^2} \cos(2 \pi \nu_\perp t + \phi) \right. \\ 
     & + &  \left. \frac{1}{3}e^{-\Delta_\parallel^2 t^2} \cos(2 \pi \nu_\parallel t +\phi) \right] \\
     & + & A_{\rm Ag} e^{-\Delta_{\rm Ag}^2 t^2} \cos(2 \pi \nu_{\rm Ag} t + \phi) \, .  
\end{eqnarray}
Here $A_{\rm s}$ and $A_{\rm Ag}$ denote the initial asymmetries of the sample and Ag backing plate contributions, respectively.
Also, $\nu_i \! = (\! \gamma_\mu/2 \pi)B_{\mu, i}$ and $\Delta_i^2 \! = \! \gamma_\mu^2 \langle {\bf B}_{\mu, i}^2 \rangle$, 
where $\gamma_\mu/2 \pi \! = \! 135.54$~MHz/T and $B_{\mu,i}$ and $\langle {\bf B}_{\mu, i}^2 \rangle$ are the local magnetic 
field and width of the field distribution at the $\mu^+$ sites ($i \! = \! \perp$, $\parallel$, and Ag), respectively. The initial
phase of the muon spin polarization is denoted by $\phi$.
As the temperature is lowered $\nu_\perp$ ($\nu_\parallel$) increases (decreases), and below $T \! \sim \! 30$~K the $\perp$ ($\parallel$) peak in the Fourier 
transform broadens and moves to the far right (left). Even so, the two sample components are observed to maintain a population ratio of 2:1 down to 1.9~K.
We note that the 6~T applied magnetic field is far below the field of 80-90~T required to close the insulating gap.\cite{Cooley:99}

\begin{figure}
\centering
\includegraphics[width=9.0cm]{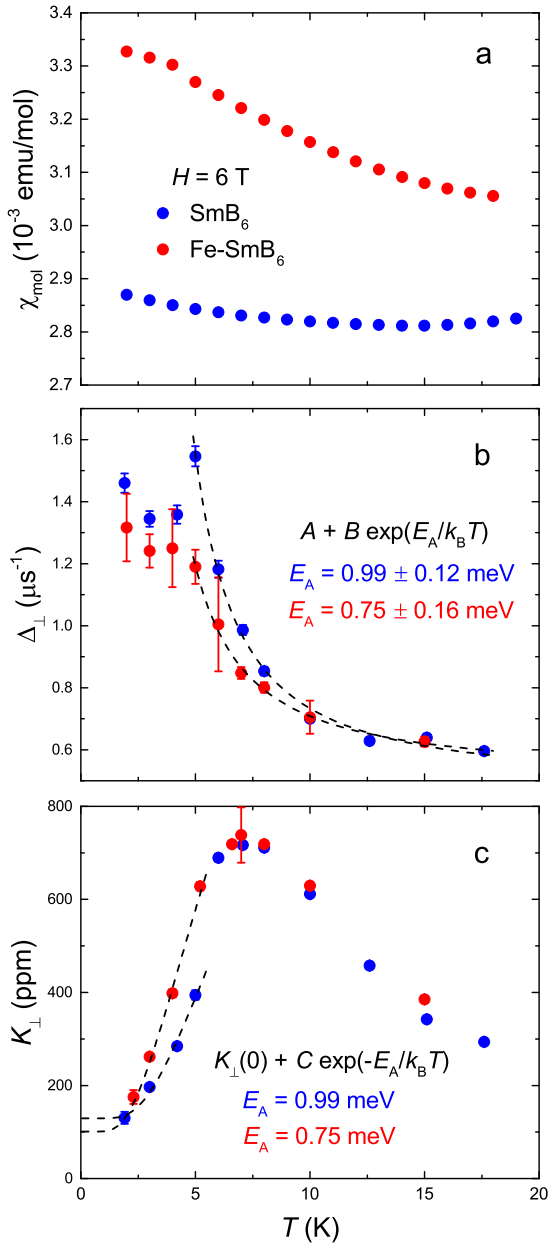}
\caption{Temperature dependences of the low-temperature {\bf a} bulk magnetic susceptibility, {\bf b} TF-$\mu$SR relaxation rate $\Delta_\perp$, and 
{\bf c} Knight shift $K_{\perp}$ in pure and 0.5~\% Fe-doped SmB$_6$ for $H \! = \! 6$~T. 
The dashed curves in {\bf b} are fits of the $\Delta_{\perp}$ data above 5~K to a thermally-activated function $A \! + \! B \exp(E_{\rm A}/k_{\rm B}T)$.
The dashed curves in {\bf c} are fits of the $K_{\perp}$ data at $T \! \leq \! 5$~K to a thermally-activated Arrhenius equation
$K_\perp(0) \! + \! C \exp(-E_{\rm A}/k_{\rm B}T)$ assuming the values of $E_{\rm A}$ from the fits in {\bf b}.}
\label{fig5}
\end{figure}

The relative frequency shift is defined as $K_{\mu, i} \! = \! (\nu_i - \nu_0)/\nu_0$, where $\nu_0 \! = \! (\gamma_\mu/2 \pi)H$.
After correcting for the demagnetization and Lorentz fields, the $\mu^+$-Knight shift at the magnetically inequivalent muon sites is 
\begin{subequations}
\begin{align}
K_{\perp}(T) & = (A_{\rm c}^{\perp} - \frac{1}{2} A_{\rm dip}) \chi_{4f}(T) + K_0^{\perp} \, , \label{eqn:Kperp} \\
K_{\parallel}(T) & = (A_{\rm c}^{\parallel} + A_{\rm dip}) \chi_{4f}(T) + K_0^\parallel \, ,
\label{eqn:Kpara}
\end{align}
\end{subequations}
where $A_{\rm c}^i$ and $A_{\rm dip}$ are the contact hyperfine and dipolar coupling constants, $\chi_{\rm 4f} \! = \! \chi_{\rm mol} \! - \! \chi_0$  
is the local $4f$ magnetic susceptibility, and $\chi_0$ and $K_0^i$ are the Pauli paramagnetic susceptibility and corresponding Knight shifts, respectively.

Figure~\ref{fig3} shows the temperature dependence of the Knight shifts $K_i$ and TF relaxation rates $\Delta_i$. In contrast to
$\chi_{\rm mol}$, the Knight shifts below 17~K do not exhibit an appreciable field dependence. We attribute the jump (dip) in $\Delta_\perp$ 
($\Delta_\parallel$) between 20 and 40~K to fit parameters playing off against each other, as $\nu_\perp \! \sim \! \nu_\parallel$ in this
temperature range.
       
Figure~\ref{fig4}a shows the Knight shift data plotted versus $\chi_{\rm mol}$ 
with temperature as an implicit parameter (a so-called Clogston-Jaccarino plot). 
We find that a plot of $K_\perp \! - \! K_\parallel$ versus $\chi_{\rm mol}$ above 110~K (not shown) is linear as expected 
from equations (\ref{eqn:Kperp}) and (\ref{eqn:Kpara}), but has a slope and intercept incompatible with $A_{\rm c}$ and $K_0$ being isotropic. 
Using the calculated value $A_{\rm dip} \! = \! 0.395$~T/$\mu_B$ for the $\mu^+$ site, linear fits of 
the Knight shifts in SmB$_6$ above 110 K to equations (\ref{eqn:Kperp}) and (\ref{eqn:Kpara}) yield 
$A_{\rm c}^{\perp} \! = \! 0.346$~T/$\mu_B$ and $A_{\rm c}^{\parallel} \! = \! -0.378$~T/$\mu_B$. 
This anisotropy can be explained by the influence of the Sm $4f^5$-shell electric quadrupole moment\cite{Delyagin:16} on the 
conduction electron spin polarization at the $\mu^+$ site, which has been observed in other compounds 
with non-spherical $f$-electron distributions.\cite{Schenck:03}

Below 110~K, the $K_i$ versus $\chi_{\rm mol}$ plots deviate from linearity (see Fig.~\ref{fig4}a). Point-contact spectroscopy\cite{Zhang:13} and 
ARPES\cite{Xu:14b} measurements on SmB$_6$ show the hybridization gap develops over a fairly wide temperature range of 30~K~$\! \lesssim T \! \lesssim \! 110$~K. 
This results in a loss of scaling between $K_i$ and $\chi_{\rm mol}$, due to a gradual reduction of both the Pauli susceptibility ($\propto \! K_0^i$)
and the electronic spin density at the $\mu^+$ sites ($\propto \! A_{\rm c}^i$). Near 30~K, the simultaneous vanishing of $K_{\perp}$ and $K_{\parallel}$ 
implies $K_0^i \! = \! 0$ and $\chi_{4f} \! = \! 0$ in equations (\ref{eqn:Kperp}) and (\ref{eqn:Kpara}).

Below $T \! \sim \! 25$~K a $\mu^+$-Knight shift reappears, which does not linearly scale with $\chi_{\rm mol}$ (Fig.~\ref{fig4}a)
and is accompanied by an increase in the TF relaxation rates with decreasing $T$ (Fig.~\ref{fig3}b). \\

\noindent {\bf DISCUSSION} \\ 
The $\mu^+$-Knight shift below 25~K is a property of the insulating bulk. In insulators and semiconductors the $\mu^+$ 
sometimes forms a bound state with an electron, known as a {\it muonium} atom (Mu).\cite{Schenck:85}
The signature of Mu in high TF is a pair of frequencies separated by the Mu hyperfine splitting and centered on the precession frequency of
the free $\mu^+$ in the applied field --- the latter being close to the $\mu^+$ precession frequency in the Ag backing plate. 
This is clearly not observed in Fig.~\ref{fig2}. Consequently, the $\mu^+$-Knight shift must still be induced by the Sm-4$f$ moments.

The lack of scaling of $K_i$ with the bulk magnetic susceptibility $\chi_{\rm mol}$ below 25~K 
could potentially arise from the charged muon significantly altering the Sm$^{3+}$ crystal electric field (CEF) level scheme and hence $\chi_{4f}$.
A significant influence of the $\mu^+$ on the local magnetic susceptibility has been identified in a few Pr$^{3+}$-ion systems.\cite{Feyerherm:95,Tashma:97}
The CEF level scheme of Sm$^{3+}$ ($4f^5$) in SmB$_6$ is similar to Ce$^{3+}$ ($4f^1$) in CeB$_6$. In both cases the spin-orbit interaction
splits the $4f$ states into $J \! = \! 5/2$ and $J \! = \! 7/2$ multiplets. The $J \! = \! 5/2$ multiplet is further split in the
cubic crystalline field into a $\Gamma_7$ doublet, and a ground-state $\Gamma_8$ quartet that has magnetic and quadrupolar moments.
The energy difference between the $\Gamma_8$ quartet and excited $\Gamma_7$ doublet is about 15~meV in SmB$_6$,\cite{Sundermann:18} 
and 46~meV in CeB$_6$,\cite{Sundermann:17} which in both compounds exceeds the Kondo energy scale (temperature). 
Thus, only modifications of the Zeeman split $\Gamma_8$ quartet are relevant in the low $T$ regime.
In CeB$_6$, which does not develop a Kondo insulating gap, $K_i$ linearly scales with $\chi_{\rm mol}$ above 10~K.
Hence it is unlikely that the $\mu^+$ induces the Knight shift observed in SmB$_6$ below 20~K. We note that
the loss of scaling between $K_i$ and $\chi_{\rm mol}$ in CeB$_6$ below 10~K is due to the development of
antiferroquadrupolar ordering,\cite{Schenck:04} which does not occur in SmB$_6$.

According to equations (\ref{eqn:Kperp}) and (\ref{eqn:Kpara}), there must be a new contact hyperfine field ${\bf B}_{\rm c}$ 
to cause the sign change in the values of $K_{\perp}$ and $K_{\parallel}$ below 25~K. As mentioned earlier,
in an insulating state this may result from the $\mu^+$ bonding to an ion that is covalently bonded to a localized magnetic electron.
A {\it super-transferred hyperfine field} at the $\mu^+$ site through a Sm-B-$\mu^+$ connection could arise from field-induced 
moments at the B sites. In CeB$_6$, field-induced magnetic moments inside or around the 
B$_6$ octahedron have been ruled out by polarized neutron diffraction,\cite{Givord:03} which is presumably 
also the case in SmB$_6$. Moreover, the formation of a B-$\mu^+$ bond is
incompatible with the $\mu^+$ site, which is $\sim \! 2$~\AA~ from the nearest B atom. 

The alternative possibility in an insulating state is that ${\bf B}_{\rm c}$
originates from direct overlap of the $\mu^+$ with the wavefunction of the localized magnetic electrons.
While the Sm-$4f$ orbitals are highly localized, the $5d$ orbitals of the nearest-neighbor Sm ions
partially overlap the $\mu^+$ site. In a spin exciton the spin polarization of the bound $5d$ electron
is coherently coupled to the localized $4f$ electrons, and a contact hyperfine field may result from
an exchange interaction between the $\mu^+$ and the extended magnetic $5d$ electrons.

The temperature dependence of the TF-$\mu$SR relaxation rate (Fig.~3b) provides evidence for a low-energy spin exciton. 
The marked increase of $\Delta_{\perp}$ and $\Delta_{\parallel}$ below 20~K corresponds to an increase in the width of the 
local field distribution, indicative of a gradual slowing down of magnetic fluctuations. 
As shown in the inset of Fig.~\ref{fig3}b, the $H \! = \! 6$~T data for $\Delta_{\perp}$ above 5~K can be fit
with a thermally-activated law: $\Delta_{\perp} \! = \! A \! + \! B \exp(E_{\rm A}/k_{\rm B} T)$, yielding
$A \! = \! 0.35 \! \pm \! 0.07$~$\mu$s$^{-1}$, $B \! = \! 0.12 \! \pm \! 0.04$~$\mu$s$^{-1}$, and $E_{\rm A} \! = \! 0.99 \! \pm \! 0.12$~meV. 
A temperature-independent contribution comes from the nuclear dipole moments and 
the spatial inhomogeneity of the applied magnetic field. The thermally-activated decrease of $\Delta_{\perp}$ 
is consistent with a rising fluctuation rate $(1/\tau)$, where $1/\tau \! \propto \! \exp(-E_{\rm A}/k_{\rm B} T)$. 
Spin excitons create fluctuating regions of AFM correlations extending over a few unit cells, which modify the local fields sensed 
by the $\mu^+$. The increase in $\Delta_i$ is explained by AFM amplitude fluctuations perpendicular to {\bf H}, producing a small temporary canted moment.
This is presumably of order $H/J_{\rm RKKY}$, where $J_{\rm RKKY}$ is the virtual 
RKKY-like magnetic exchange interaction between the $4f$ moments in the theory of Riseborough.\cite{Riseborough:00,Riseborough:03}
The nearly constant difference between $\Delta_i$ at 6~T and 4~T above 6~K is primarily caused
by a difference in the inhomogeneity of the applied field. Below 5-6~K, however, $\Delta_i$ saturates
and exhibits an intrinsic increase with $H$. The saturation indicates that the average fluctuation period of
the spin excitons ($\tau$) has become large with respect to the muon time window, such that $\Delta_i$ is no longer
significantly affected by fluctuations. The corresponding field-dependence of $\Delta_i$ below 5-6~K reflects 
an inhomogeneous broadening of the local field distribution.

Figure~\ref{fig4}b shows a strong temperature dependence of 
$K_{\perp}/K_{\parallel}$ at $6 \! \lesssim \! T \! \lesssim \! 20$~K, which with $K_0^i \! = \! 0$ reflects the behavior of 
$(A_{\rm c}^{\perp} \! - \! A_{\rm dip}/2)/(A_{\rm c}^{\parallel} \! + \! A_{\rm dip})$.
Changes in the lattice parameter\cite{Tarascon:80,Sirota:98,Mandrus:02} below 20~K are too small to cause an appreciable 
change in $A_{\rm dip}$. The remaining possibility is that $A_{\rm c}^{\perp}$ and $A_{\rm c}^{\parallel}$ change with
decreasing temperature. The anisotropy of $A_{\rm c}$ above 110~K indicates coupling of the non-spherical $4f$-electron 
distribution to the field-induced Sm magnetic moments. Hence, the strong temperature dependence of 
$A_{\rm c}^{\perp}$ and $A_{\rm c}^{\parallel}$ below 20~K is likely due a rotation of the $4f$-electron 
distribution with the canted moment induced by AFM fluctuations. We note that the values of $A_{\rm c}^i$ are
dependent on the overlap integral of the non-spherical $4f$-electron distribution and the $5d$ electrons with
wavefunctions that overlap the $\mu^+$ site. Below 6~K there is an abrupt increase in
$K_{\perp}/K_{\parallel}$ to a value comparable to that at 25~K. This suggests that 
the $4f$-electron distribution returns to being more closely aligned with the applied field.  

As shown in Fig.~\ref{fig3}a, there is an abrupt decrease (increase) in $K_\perp$ ($K_\parallel$) below 5-6~K.
According to equations (\ref{eqn:Kperp}) and (\ref{eqn:Kpara}), these simultaneous behaviors cannot be 
explained by a rapid change in $\chi_{4f}(T)$. Instead these behaviors appear to reflect the temperature
dependence of $K_{\perp}/K_{\parallel}$ (and hence $A_{\rm c}^{\perp}$ and $A_{\rm c}^{\parallel}$) that is
apparent at $T \! \leq \! 5$~K in Fig.~\ref{fig4}b. As shown in Fig.~\ref{fig4}c, the temperature dependence of
$K_\perp$ for $T \! \leq \! 5$~K is well fit with a thermally-activated Arrhenius equation: 
$K_\perp(T) \! = \! K_\perp(0) \! + \! C \exp(-E_{\rm A}/k_{\rm B} T$ that assumes the same value 
$E_{\rm A} \! = \! 0.99$~meV obtained from the fit of the temperature dependence of $\Delta_\perp$.

The dynamic relaxation rate $\lambda_{\rm ZF}$ observed by zero-field (ZF) $\mu$SR develops below 20-25~K,\cite{Biswas:14}
exhibits a short anomalous peak near 4~K and subsequently saturates. Our findings here suggest that the saturation is 
due to the freezing out of a bulk spin exciton of much lower energy than that observed by INS, giving way to 
AFM quantum spin fluctuations.  We note that the peak in $\lambda_{\rm ZF}(T)$ near 4~K 
vanishes with the addition of a 0.5~\% Fe impurity.\cite{Akintola:17} This is likely due to the predicted adverse 
effect of impurities on a fully developed spin exciton.\cite{Riseborough:03}
We have carried out similar TF-$\mu$SR measurements on the 0.5~\% Fe-doped sample. 
Figure~\ref{fig5} shows a comparison of the low-temperature results with those for the pure compound. 
While the bulk magnetic susceptibility is significantly modified by the Fe impurities, the changes to the 
temperature dependences of $\Delta_\perp$ and $K_\perp$ are more subtle. A fit of the $\Delta_\perp$ versus $T$ data for the 
Fe-doped sample above 5~K to the thermally-activated law described earlier yields $E_{\rm A} \! = \! 0.75 \! \pm \! 0.16$~meV.
The smaller activation energy compared to pure SmB$_6$ is consistent with the expected impurity-induced
broadening and decrease of the binding energy of the spin exciton.\cite{Riseborough:03}     

As shown in Fig.~\ref{fig4}c, the saturation of the electrical 
sheet resistance $R_{\rm s}$ occurs below 4~K, where $K_\perp$ is greatly reduced. Thus
the resistivity plateau apparently occurs when spin exciton scattering of the metallic surface states becomes negligible.
Our findings support the theoretical prediction of a low-energy spin exciton ($\lesssim \! 1$~meV) in SmB$_6$, which
has been argued to account for certain low-temperature thermodynamic and transport anomalies.\cite{Knolle:17} \\

\noindent {\bf METHODS} \\
Samples and experimental technique\\
\footnotesize{The growth and characterization of the crystals were previously reported.\cite{Akintola:17} The high TF $\mu^+$-Knight shift measurements were performed on a $5 \! \times \! 5$~mm$^2$ mosaic of single crystals aligned with the $c$-axis parallel to the applied field and mounted on a pure Ag backing plate. 
The measurements utilized a He-gas flow cryostat and the so-called ``NuTime" spectrometer at TRIUMF in Vancouver, Canada. All of the high TF-$\mu$SR measurements were recorded with the initial muon spin polarization {\bf P}(0) perpendicular to the external magnetic field {\bf H}, which was applied parallel to the muon beam momentum. Figure S1 shows a schematic of a custom sample holder designed for the $\mu^+$-Knight shift measurements.

The value of the muon spin precession frequency in the applied magnetic field $H$ alone, $\nu_0 \! = \! (\gamma_\mu/2 \pi)H$, was accurately determined by 
first simultaneously recording the TF-$\mu$SR signal in a 99.998~\% pure Ag mask located upstream of the sample. To correct for the difference in the 
external field at the Ag mask and sample locations, TF-$\mu$SR measurements were also simultaneously performed on the Ag mask with Ag in place of 
SmB$_6$ at the sample location.} \\

\noindent Data availability \\
\footnotesize{All relevant data are available upon request from the corresponding authors.} \\

\noindent {\bf ACKNOWLEDGEMENTS} \\
This work was performed at TRIUMF, the University of Maryland (UofM) and Simon Fraser University (SFU).
J.E.S. acknowledges support from NSERC of Canada. J.P. acknowledges support from AFOSR  
through Grant No. FA9550-14-1-0332 and the Gordon and Betty Moore Foundation's EPiQS Initiative through Grant No. GBMF4419.
The authors wish to thank E. Mun, J. S. Dodge and P. S. Riseborough for informative discussions. \\

\noindent {\bf AUTHOR CONTRIBUTIONS} \\
S.R.S, X.F.W., J.P., grew and characterized the sample. K.A. performed the bulk magnetic susceptibility measurements.
K.A., A.P., S.R.D., A.C.Y.F., M.P., S.R.S., and J.E.S. performed the high TF-$\mu$SR measurements at TRIUMF. 
K.A. and J.E.S. carried out the data analysis. K.A. and J.E.S. wrote the manuscript with input from all co-authors. \\

\noindent {\bf ADDITIONAL INFORMATION} \\

\noindent \footnotesize{{\bf Supplementary information} accompanies the paper.} \\

\noindent \footnotesize{{\bf Competing interests:} The authors declare no competing interests.}

\end{document}